

\input jnl.tex

\def\de{{\delta}}
\def\eps{{\epsilon}}

\def\th{{\theta}}

\def\si{{\sigma}}

\def\cG{{\cal G}}

\def\cM{{\cal M}}
\def\gtwid{\raise.3ex\hbox{$>$\kern-.75em\lower1ex\hbox{$\sim$}}}
\def\ltwid{\raise.3ex\hbox{$<$\kern-.75em\lower1ex\hbox{$\sim$}}}
\def\vev#1{{\left\langle{#1}\right\rangle}}

\rightline{NSF-ITP-93-136}

\title Universal Spectral Correlation
between Hamiltonians with Disorder

\vskip .2in

\author E. Br\'ezin$^1$ and A. Zee$^2$

\affil
$^1$Laboratoire de Physique Th\'eorique
\'Ecole Normale Sup\'erieure
24 Rue Lhomond
75231 Paris, France

\affil
$^2$ Institute for Theoretical Physics
University of California
Santa Barbara, CA 93106-4030 USA

\abstract
{We study the correlation between the energy spectra of
two disordered
Hamiltonians of the form $H_a = H_{0a} + s_a\varphi$
 (a=1,2) with $H_{0a}$ and $\varphi$ drawn from random
distributions. We
calculate this correlation function explicitly and show that
it has a simple
universal form for a broad class of random distributions.}

\vskip .5in

In a series of recent papers [\cite{BZ1, BZ2, BZ3}]
(hereafter to
be referred to as I, II, and III respectively) we studied the
theory of random matrices [\cite{WIG,POR,MEH}] and
discovered a remarkably simple universality in the
correlation
between
energy eigenvalues of random Hamiltonians. A brief
review and
summary of our work may be found in III.
In particular, we find that for
a large class of Hamiltonians, the smoothed connected
correlation (to be defined below) between the density of
energy
eigenvalues at energies $\mu$ and $\nu$ has the universal
form
$$
\rho^{{\rm smooth}}_c (\mu,\,\nu)  =
{-1\over2(N\pi a)^2 } {1\over(x-y)^2}
\left\{(1-xy)\over[(1-x^2)(1-y^2)]^{1/2}\right\}
. \eqno(universal)
$$
(where we have centered the spectrum at the origin.)
Here $a$ denotes the endpoint of the spectrum and we
have introduced the
scaled
variables $x=\mu/a$ and
$y=\nu/a$. We see that up to a factor of $1/a^2$, $\rho^{{\rm
smooth}}_c$ is equal
to a universal
function of $x$ and $y$.
The
only dependence on the random distribution of
Hamiltonians
appears through $a$. This result is all the more remarkable
since,
by way of contrast, the density $\rho(\mu)$ is
completely non-universal. This universality has also been
obtained recently by  Beenakker [\cite{bee}] using another
method. For $\mu$ and $\nu$ far from the endpoints of the spectrum,
we
see that in \(universal) the factor in the curly bracket is essentially
equal to
one, and so the form the universal correlation function basically
says that
there is an algebraic long-distance correlation falling like $1/(x-
y)^2$.
In III, we developed an efficient diagrammatic
method
to calculate the
density of eigenvalues and the correlation between them.

Meanwhile, in the recent literature
Altschuler and collaborators [\cite{SSA}] have
considered the
correlation between the energy spectra of two
Hamiltonians
$H_1$ and $H_2$.    In particular, Simons and Altschuler
considered the problem of a
single electron moving in a ring threaded by a
magnetic flux and
with
the electron scattering on impurities in the ring. The
magnetic
flux
is then changed to some other value. Here $H_a = H_{0a}
+
\varphi$ for $a = 1, 2$, where $H_{0a}$ are
deterministic, representing the
interaction of the electron with the magnetic flux,
while
$\varphi$ is random, representing the interaction of the
electron with the
impurities. The correlation
between
the spectra of $H_1$ and $H_2$
is of interest in the physics of mesoscopic
systems. In III, we determined this correlation for a certain
class of distribution of $\varphi$.

More generally, we may consider the correlation between
the
spectra of two Hamiltonians $H_a = H_{0a} + s_a\varphi$
for $a =
1, 2$, where $s_a$ represent the strength of the disorder.
Simons, Altschuler, and Lee [\cite{SSA}] and more
recently
Narayan and Shastry [\cite{shastry}] have considered this
problem for
the simpler case of $H_{01}=H_{02} \equiv H_0$. In this
paper, we
would like to demonstrate the power of the diagrammatic
approach discussed in III by calculating the correlation
between
the spectra  of $H_1$ and $H_2$ for the more general case
in which
$H_{01}\not=H_{02}$. Furthermore, instead of obtaining
an implicit integral
representation for the correlation, we will determine it
explicitly.

It may be useful to clarify the difference [\cite{diff}] between our work and
the
work
of Altschuler et al. They focus [\cite{SSA,shastry}] on the
correlation
function $\rho_c(\mu,\nu)$ for $\mu - \nu$ small, of order $1/N$,
(that is, for ``distances" much smaller than the width of the
spectrum), and
indeed often for $\mu$ and $\nu$ both small. In III and in this work
(and
in the recent work of Beenakker [\cite{bee}]) we look at the
correlation function for
$\mu - \nu$ of order $N^0$. In other words, we look at long distance
correlation, while Altshuler and collaborators focus on short
distance
correlation. In fact, in I, by determining the large-index behavior of the
relevant orthogonal polynomials, we were able to calculate explicitly the
$complete$ correlation
function over all ``distance" scales, from short distances through
intermediate distances to long distances. The correlation function discussed
in the literature appears as a limiting case of the correlation function
found in I. The work in I thus provides a bridge between the short distance
discussion of Altschuler et al and the long distance discussion we gave in II
and III.

Let us begin by recalling some basic definitions and
formulas.
Consider an ensemble of $N$ by $N$ hermitean matrices
$\varphi$ defined
by a
probability distribution
$
P(\varphi)
$
and normalized by
$\int d
\varphi
P(\varphi)=1$. The large N limit will always be
understood.
Define as usual the Green's functions
$$
\eqalignno{
G(z)&\equiv\vev{{1\over N}tr{1\over z-
\varphi}}&(1.3)\cr
G(z,w)_c
&\equiv\vev{{1\over N}tr{1\over z-\varphi}{1\over
N}tr{1\over
w-\varphi} }_c&
(1.4)
\cr}
$$
where
$
\vev{0 (\varphi)} \equiv \int d\varphi\  0(\varphi)P(\varphi)
$
and the subscript $c$ indicates the connected Green's
function.
The density of eigenvalues is then given by
$$
\rho(\mu)=\vev{{1\over N}tr\de(\mu-\varphi)}={-
1\over\pi}{\rm Im}
 G(\mu+i\eps)
\eqno(1.6)
$$
and the correlation between eigenvalues, by
$$
\eqalign{
\rho_c(\mu,\nu)&=\vev{{1\over N}tr\de(\mu-
\varphi){1\over
N}tr\de(\nu-
\varphi)}_c\cr
&=(-1 /4\pi^2)(G_c(++)+G_c(--)-G_c(+-)-G_c(-
+))\cr}\eqno(1.7)
$$
with the obvious notation
$
G_c(\pm,\pm)\equiv G_c(\mu\pm i\eps,\nu\pm
i\de)
$
(signs uncorrelated).  It is worth remarking that there are two
limits involved: $N \to\infty$ and $\eps, \delta \to 0$. As explained
in III, we obtain the full correlation function (from which we can
read off the correlation at all distances) by letting $\eps, \delta \to
0$ first (the limit we took in I), and the smoothed correlation
function by taking $N \to\infty$ first. In this paper, we will look at
the smoothed correlation.

For
applications to disordered system $\varphi$ is to be
thought
of as the
Hamiltonian. Its eigenvalues then describe the energy
levels of
the system.
In
some applications, $\varphi$ is related to the transmission
matrix [\cite{PICH}]. From the density of and correlation
between its
eigenvalues we learn
about the fluctuation of the
conductance in disordered metals.

In this work, we wish to study the correlation between the
spectra of two
different Hamiltonians $H_1$ and $H_2$. Evidently, we
can determine this
quantity by studying the generalized (connected) Green's
function
$$
\cG(z,w)_c
\equiv\vev{{1\over N}tr{1\over z-\varphi_1}{1\over
N}tr{1\over
w-\varphi_2} }_c     \eqno(crossG)
$$
where to conform to field theoretic notation we have
renamed
$H_a$ as $\varphi_a$, two ``matrix scalar fields." Note
that our starting point
appears to correspond to one of the concluding results of
\Ref{shastry}. (Cf.
eq. (16) there.)

For ease of presentation we will begin with
the Gaussian distribution
$$
P(H_{01}, H_{02}, \varphi)={1\over Z}e^{-{N\over 2}
tr [m^2_1 H^2_{01}
+ m^2_2 H^2_{02} +
m^2 \varphi^2 + \mu^2 (H_{01}- H_{02})^2]
}.\eqno(distribution)
$$
Note that the factors of $N$ are
chosen in
our definitions such that in the large $N$ limit the density
of eigenvalues has
a
finite
(\ie of order
$N^0$) width. For $\mu^2 \rightarrow \infty$, we recover
the case
$H_{01}=H_{02}$
considered earlier [\cite{SSA,shastry,BZ3}]. In the
opposite limit
$\mu^2=0$, the two Hamiltonians $H_{01}$ and
$H_{02}$ are completely
independent. In
evaluating
$\cG(z,w)_c$, we see that, since the integrand depends
only on
$\varphi_a$, after using the definition $\varphi_a =
H_{0a} +
s_a\varphi$ to eliminate $H_{0a}$ and integrating over
$\varphi$,  we can
effectively use $P(\varphi)={1\over Z'} e^{-
{N\over 2} tr (\varphi_a
M^2_{ab}\varphi_b)}$ with the 2-by-2 symmetric matrix
$M^2$ defined by
$$
M^2_{11}=(m^2_1+\mu^2)-\cM^{-
2}[(m^2_1+\mu^2)s_1-\mu^2
s_2]^2,\eqno(m1)
$$
and similarly for $M^2_{22}$, and
$$
M^2_{12}=-\mu^2-\cM^{-2}[(m^2_1+\mu^2)s_1-\mu^2
s_2][(m^2_2+\mu^2)s_2-\mu^2 s_1],\eqno(m2)
 $$
with
$$
\cM^2  \equiv m^2+\sum((m^2_a+\mu^2)s^2_a-
2s_1s_2\mu^2.\eqno(m3)
$$
Let us now expand
$$
\cG_c(z,w)={1\over N^2}
\sum^\infty_{m=0}\sum^\infty_{n=0}{1\over
z^{m+1}w^{n+1}}
\vev{tr\varphi^m_1(t)tr\varphi^n_2(0)}_c     \eqno(expand)
$$
Diagrammatically, we may borrow the
terminology of large N
QCD [\cite{thoo}]
and describe the expression for
 $\cG_c(z,w)$ as two separate quark loops,
of type $z$ and type $w$ respectively, interacting by
emitting
and absorbing gluons.  A quark of type $z$ can only emit and
absorb gluons  of
``type 1" and a quark of type $w$ and only emit and
absorb
gluons of ``type
2." With a Gaussian distribution for $\varphi_a, a =1,2$,
we can readily
``Wick-contract" \(expand).
The inverse gluon propagator is given by
$$
\vev{\varphi_{aij} \varphi_{bkl}(0)} =
{1\over N}\de_{il}\de_{jk}\sigma^2_{ab}\eqno(glueprop)
$$
where the 2-by-2 matrix $\sigma^2 \equiv M^{-2}$. (For the sake
of notational
simplicity, we will define $\sigma^2 = (\sigma^2)_{11}$,
$\tau^2 =
(\sigma^2)_{22}$, and
$\rho^2 = (\sigma^2)_{12}$.) Thus, the gluon is represented by a
double
line while a quark  is represented by a single line. This convention
greatly
facilitates counting the powers of $N$.

 Let us begin by ignoring
contractions
within
the same trace (in which case $m$ and $n$ are required to
be
equal). In the
large $N$ limit, the dominant graphs are given essentially by ``ladder
graphs"
(with one crossing) which immediately sum to
$$
N^2 \cG_c(z,w) = {\rho^2\over(z w)^2}  {1\over(1-
{\rho^2\over z w})^2}
\eqno(ladder)
$$

We next include Wick-contractions within the same trace
in
$\vev{tr\varphi^m_1
tr\varphi^n_2}$. Graphically these contractions
correspond to decorating the ladder graphs by vertex
and self
energy corrections. Summing the vertex corrections, we
see that we have to
multiply the expression in \(ladder) by two factors, the
factor
$(1-{\sigma^2\over z^2})^{-2}$
and a similar factor with $z$ and $\sigma$ replaced by
$w$ and $\tau$
respectively. Finally, summing the self energy corrections
we see that the
bare quark propagator $1/ z$ gets dressed to
$$
G(z, \sigma) ={1\over2\si^2}(z -  \sqrt{z^2-4{\sigma^2}})
\eqno(green)
$$
(and
similarly for $1 / w$ of course with $\sigma$ replaced by
$\tau$.) Note that the bare propagators are
recovered in the limit
$\sigma,\tau \rightarrow 0$. Putting these various factors
together, we
obtain the remarkably
compact
result
$$
N^2\cG_c(z,w)= {\rho^2\over(1-
\rho^2G(z,\sigma)G(w,\tau))^2}
\left[{G^2(z,\sigma)\over1-
\sigma^2 G^2(z,\sigma)}\right]
\left[{G^2(w,\tau)\over1-\tau^2 G^2(w,\tau)}\right]
\eqno(connected2point)
$$
As indicated in III, we find it convenient to introduce
angular variables:
$
\sin \theta \equiv \mu / 2\sigma
$
and
$
\sin \phi \equiv \nu / 2\tau.
$
 As $\theta$ and $\phi$
vary
from $-\pi/ 2$ to $\pi/ 2$, $\mu$ and
$\nu$ vary over of the width of their
respective spectra. These angular variables are the natural scaling
variables
to use in this class of problems dealing with eigenvalues of random
matrices.
{}From \(green), we have
$$
G(\mu+i\epsilon, \sigma) = -i \eta e^{i \eta \theta}/\sigma
\eqno(cutG1)
$$
where $\eta$ = the sign of $\epsilon$ and
$$
G(\nu+i\delta, \tau) = -i \xi e^{i \xi \phi} /\tau
\eqno(cutG2)
$$
with $\xi$ = the sign of $\delta$. A straightforward
computation of the appropriate generalization of \(1.7)
then yields
$$
\eqalign{
& - 16\pi^2 N^2\rho_c(\mu,\,\nu) \cr
& = {1\over\sigma\tau\cos \th\cos\phi}
\left\{{1+ {\rm ch\ u}\cos(\th+\phi)\over[{\rm ch\
u}+\cos(\th+\phi)]^2} +
{1- {\rm ch\ u}\cos(\th-\phi)\over[{\rm ch\ u}-\cos(\th-
\phi)]^2}\right\}
.\cr} \eqno(correlation)
$$
Here we have defined
$$
u \equiv \log (\sigma\tau/\rho^2) = {1\over
2}\log[M^2_{11}M^2_{22}/(M^2_{12})^2],\eqno(defineu
)
$$
a rather involved function of the parameters $m^2_1$,
$m^2_2$,
$m^2$, and $\mu^2$ that appear in the distribution
\(distribution). Perhaps
remarkably, aside from an overall factor of $\sigma\tau$
the correlation
function, when expressed in terms of the appropriately
scaled variables
$\theta$ and $\phi$ depends on the probability distribution
only through
this particular combination. This represents already a form
of universality or
scaling.

The reader familiar with III would recognize that this is the
same expression
we obtained there (see eq. (2.18) in \Ref{BZ3}) for an
apparently totally different
problem, with a totally different definition of $u$. In III
we studied the
time-dependent correlation between the eigenvalues of
time-dependent
matrices $\varphi(t)$ taken from a probability distribution
defined by
$$
P(\varphi)={1\over Z''} \exp-\int_{-T}^T dt\  Tr
\left[\varphi K({d\over dt} )
\varphi + V(\varphi)\right] \eqno(timedep)
$$
with K any reasonable
function and with
$T\to\infty$. Here ``time" may correspond to some
external
parameter we
are allowed to vary. For this problem, we defined $e^{-
u(t)}$ to be $\int
{d\omega\over2\pi}e^{i\omega t} (1/K(\omega))$.
For the problem
considered in this paper, in contrast, $e^{-u}$ is defined
simply in terms of
the widths of the distributions from which various random
Hamiltonians are
drawn. Evidently, we have in this way found a ``mapping"
between two apparently unrelated problems involving
disordered
Hamiltonians. A particularly simple case occurs when
$K({d\over dt} )=
-{d^2\over dt^2}$
in which case $u(t)$ is essentially time. It is in this sense
that loosely, one
may speak of $u$ as defined in \(defineu) as ``time," even
though there is no
notion of time as such in the problem studied in this paper.

Various limits can now be studied. In the large ``time"
limit $u \rightarrow
\infty$, we obtain
$$
4\pi^2N^2\rho_c(\mu,\,\nu)\to e^{-u}\tan\th\tan\phi
\eqno(longtime)
$$
For $\theta=\phi$ (note that this does not mean $\mu
= \nu$ when $\sigma$ is not equal to $\tau$) and small ``time" $u$
we find
$$
8\pi^2N^2\rho_c(\mu,\nu)={1\over \sigma\tau\cos^2\th} {1\over
u^2}
+\dots
\eqno(samepoint)
$$
For $\theta \neq \phi$ and $u << \theta - \phi$ we obtain
$$
 -16\pi^2 N^2\rho_c(\mu,\,\nu)={1\over\sigma\tau\cos\th\cos\phi}
\left\{ {1\over1+\cos(\th+\phi)} \left[1-{1-
{1\over2}\cos(\th+\phi) \over
       1+\cos(\th+\phi)}u^2\right] +(\phi\to-\phi+\pi) \right\}
\eqno(smalltime)
$$

We can also study various regions of parameter space, of
course. Consider
(A) $\mu \to \infty$ so that
$H_{01}=H_{02}=H_0$ in which
case it is convenient to define $m^2_0 =
m^2_1 + m^2_2$. Case (A1): with $m_0\to \infty$ (so that
$H_0 = 0$) and
$s_1=s_2$ we recover the universality (universal) found in
I and in
\Ref{bee}. Case (A2): with $m_0 >> m$ so that the
disorder is large, we find
that $e^u = \sigma\tau /{\rho^2}=1$ and thus the same
universal correlation as
in case (A1) suitably scaled by an overall factor of
$\sigma\tau$, but here
there
is no requirement
on $s_a$. Case (A3): with $s_2$ equal to zero, so that we
are studying the
correlation between the spectra of a ``bare" Hamiltonian
$H_0$ and a
Hamiltonian disturbed by an external perturbation $H_0 +
s_1
\varphi$, we find
that $e^u = 1 + m^2_0 s^2_1 /m^2$. Note that no general
statement on how
``time" $t$ maps onto the strength $s_1$ of the external
perturbation can be
made, but in the simplest case of identifying ``time"
mentioned above we
have
 $t \propto \log s_1$ for large strength of the perturbation.

Another interesting class (B) is defined by the limit
$\mu\to 0$ so that
$H_{01}$ and $H_{02}$ are not correlated at all. After
some drastic
simplification, we find $\sigma^2=m^{-2}_1 + s^2_1 m^{-
2}$, $\tau^2=m^{-2}_2
+ s^2_2 m^{-2}$, and $\rho^2=s_1 s_2 m^{-2}$. As a
check on the formalism,
we can consider some special cases. For instance, with
$m_a \to \infty, a
=1,2$, $H_{01}$ and $H_{02}$ are set to zero, in which
case we recover
essentially our previous result \(universal). As another
example, consider
the analog of case (A3): with $s_2=0$ we obtain
$u\to\infty$ and as
expected $\rho_c =0$ since now $H_1$ and $H_2$ are no
longer ``linked" by
$\varphi$.

All these results are derived with the Gaussian distribution
\(distribution). We now
remark on how these results may hold for more general
distribution. In III
we identified two classes of random matrix distribution
which we refer to as
the Wigner class and the trace class. In the Wigner class,
we consider an
ensemble of matrices whose matrix elements are
independently distributed
according to some probability distribution (the same
distribution for all the
elements.) Let us focus on the example in which the
probability of the
distribution
matrix
element $\varphi_{ij}$ is given by
$$
P(\varphi_{ij}) \propto e^{-N^2 {(|\varphi_{ij}|^2-
{v^2\over
N^2})^2 } }
.\eqno(wigner)
$$ (Here $\varphi$ represents any of the matrices in
\(distribution)
generically.) It is easy to see that in the Feynman diagrams involving
the
quartic interaction $\sim N^2|\varphi_{ij}|^4$ the indices $i$ and $j$
are
tied together so to speak and thus these graphs are suppressed by a
power
of $N^{-1}$  relative to the graphs in which this interaction vertex
does
not appear. Reasoning
along
this line, we
see immediately that the results of this paper have a generality far
beyond the Gaussian
distribution. Our results are universal. This universality is much
harder to
prove for
the so-called trace
class in which the generic random matrix $\varphi$ is
taken from a
distribution of the form $P(\varphi)={1\over Z}e^{-N tr
V(\varphi)}$. We
conjecture however that our results may also hold for
random Hamiltonians
defined with this class of probability distribution. In II, we
had shown that
the universal correlation in \(universal) indeed holds for
this
trace class and
for a much broader class generalizing the trace class.
In III, we also calculated the correlation when  $H_{01}$
and $H_{02}$ are
set to equal deterministic, rather than random,
Hamiltonians $W_1$ and $W_2$ say. This case may
be considered in the present context by adding to the
logarithm of the
distribution in \(distribution) terms like $-\alpha^2_a
(H_{0a}-W_a)^2$ and
taking $\alpha_a \to \infty$.
The gluons $\varphi_1$ and $\varphi_2$ are now coupled
also to external
sources.

In conclusion, we have extended the universality in
\(universal) to a
much broader class of problems involving random
Hamiltonians. Expressed
in terms of suitably scaled variables, the correlation
between energy
eigenvalues is universal, certainly in the Wigner class, and
most likely in the
trace class as well. The case $H_{01} = H_{02}$
discussed previously in the
literature can be recovered as a special case within our
discussion.

\head{Acknowledgement}

 This work was supported in part by the National Science
Foundation under Grant No. PHY89-04035 and the Institut
Universitaire de France.




\references

\refis{BZ1} E. Br\'ezin and A. Zee, \np 402(FS), 613, 1993.

\refis{BZ2} E. Br\'ezin and A. Zee, {\sl Compt.\ Rend.\
Acad.\
Sci.\/}, 317, 735, 1993.

\refis{diff} One of us (AZ) thanks B. Altschuler for
 a discussion clarifying this
difference.

\refis{BZ3} E. Br\'ezin  and A. Zee, ``Correlation
functions in disordered
systems," SBITP-preprint 1993.

\refis{WIG} E. Wigner, {\sl Can.\ Math.\ Congr.\ Proc.\/}
p.174
(University of
Toronto Press) and other papers reprinted in Porter, op.
cit.

\refis{POR} C.E. Porter, {\it Statistical\ Theories\ of\
Spectra:\ \
Fluctuations\/}
(Academic Press, New York, 1965).

\refis{MEH} M.L. Mehta, {\it Random\ Matrices\/}
(Academic
Press, New
York,
1991).

\refis{shastry} O. Narayan and B. S. Shastry, \prl 71,
2106, 1993.

\refis{bee} C.W.J. Beenakker, Institut Lorentz preprint
(1993) cond-mat/9310010; for earlier work, see C.W.J.
Beenakker, \prl 70, 1155, 1993; \pr
B47, 15763, 1993.

\refis{SSA} B.D. Simons and B.L. Altschuler, \prl 70,
4063, 1993; B. D. Simons,
P. A. Lee, and B. L. Altschuler, \prl 70, 4122, 1993.





\refis{PICH} J.-L. Pichard, in {\it Quantum\ Coherence \
in \
Mesoscopic\ \
Systems\/} (NATO ASI series, Plenum, New York 1991)
ed. by B.
Kramer.

\refis{thoo} G. 't Hooft, \np B 72, 461, 1974.

\endreferences

\end